\documentclass[conference, twocolumn, 10pt, final, letterpaper]{IEEEtran}
\IEEEoverridecommandlockouts
\usepackage{cite}
\usepackage{amsmath,amssymb,amsfonts}
\usepackage{algorithmic}
\usepackage{graphicx}
\usepackage{textcomp}
\usepackage{xcolor}
\def\BibTeX{{\rm B\kern-.05em{\sc i\kern-.025em b}\kern-.08em
    T\kern-.1667em\lower.7ex\hbox{E}\kern-.125emX}}

\usepackage[hidelinks]{hyperref}
\usepackage{csquotes}
\usepackage{tikz}

\newcommand\copyrighttext{%
  \footnotesize \textcopyright 2023 IEEE. Personal use of this material is permitted. Permission from IEEE must be obtained for all other uses, in any current or future media, including reprinting/republishing this material for advertising or promotional purposes, creating new collective works, for resale or redistribution to servers or lists, or reuse of any copyrighted component of this work in other works. DOI: \href{https://doi.org/10.1109/BigData59044.2023.10386708}{10.1109/BigData59044.2023.10386708}
}
\newcommand\copyrightnotice{%
\begin{tikzpicture}[remember picture,overlay]
\node[anchor=south,yshift=10pt] at (current page.south) {\fbox{\parbox{\dimexpr\textwidth-\fboxsep-\fboxrule\relax}{\copyrighttext}}};
\end{tikzpicture}%
}
\makeatletter
\def\ps@headings{% default to standard twoside headers, no footers
\def\@oddhead{\hbox{}\@IEEEheaderstyle\rightmark\hfil\thepage}\relax
\def\@evenhead{\@IEEEheaderstyle\thepage\hfil\leftmark\hbox{}}\relax
\let\@oddfoot\@empty
\let\@evenfoot\@empty
}
\def\ps@IEEEtitlepagestyle{% default title page headers, no footers
\def\@oddhead{\hbox{}\@IEEEheaderstyle\leftmark\hfil\thepage}\relax
\def\@evenhead{\@IEEEheaderstyle\thepage\hfil\leftmark\hbox{}}\relax
\def\@oddfoot{\copyrightnotice}
\def\@evenfoot{\copyrightnotice}
}
\makeatother
\begin{document}

\title{A Modular Approach to Automatic Cyber Threat Attribution using Opinion Pools
% \thanks{Dutch Research Council (NWO), Project NWA.1215.18.003}
}

\author{\IEEEauthorblockN{Koen T. W. Teuwen}
\IEEEauthorblockA{\textit{Department of Mathematics and Computer Science} \\
\textit{Eindhoven University of Technology}\\
Eindhoven, The Netherlands \\
k.t.w.teuwen@tue.nl}
}

% \IEEEpubid{979-8-3503-2445-7/23/\$31.00 \copyright 2023 European Union}
% \IEEEpubid{\makebox[\columnwidth]{979-8-3503-2445-7/23/\$31.00 \copyright2023 IEEE \hfill}\hspace{\columnsep}\makebox[\columnwidth]{ }}

\maketitle

% \IEEEpubidadjcol

\begin{abstract}
Cyber threat attribution can play an important role in increasing resilience against digital threats. Recent research focuses on automating the threat attribution process and on integrating it with other efforts, such as threat hunting. To support increasing automation of the cyber threat attribution process, this paper proposes a modular architecture as an alternative to current monolithic automated approaches. The modular architecture can utilize opinion pools to combine the output of concrete attributors. The proposed solution increases the tractability of the threat attribution problem and offers increased usability and interpretability, as opposed to monolithic alternatives. In addition, a Pairing Aggregator is proposed as an aggregation method that forms pairs of attributors based on distinct features to produce intermediary results before finally producing a single Probability Mass Function (PMF) as output. The Pairing Aggregator sequentially applies both the logarithmic opinion pool and the linear opinion pool. An experimental validation suggests that the modular approach does not result in decreased performance and can even enhance precision and recall compared to monolithic alternatives. The results also suggest that the Pairing Aggregator can improve precision over the linear and logarithmic opinion pools. Furthermore, the improved $k$-accuracy in the experiment suggests that forensic experts can leverage the resulting PMF during their manual attribution processes to enhance their efficiency.
\end{abstract}

% Depending on how vigilant their paper processor is, IEEE may ask for these in your final paper, but we've heard about these amazing inventions called search engines that are able to index every word in your paper, so no need to include them in your submission unless you really want to.

\begin{IEEEkeywords}
Cyber Threat Attribution, Modular Architecture, Opinion Pools, Cyber Threat Intelligence, Digital Forensics
\end{IEEEkeywords}

\section{Introduction}
\label{sec:introduction}
Digital systems are omnipresent in our daily life. However, threat actors exist that try to exploit these systems for their own needs or simply to disrupt the activities of legitimate users. In addition to prevention of security incidents, organizations may also invest in incident detection and response \cite{schneier-defense-in-depth}. If an incident occurs despite preventive measures, digital forensic experts can attempt to attribute these incidents to threat actors. This process is sometimes referred to as cyber threat attribution. The most evident use case for attribution is that it may result in the prosecution of actors responsible for incidents. Attribution can also contribute to establishing a proactive defense by providing insights into the potential future steps of an attacker using Cyber Threat Intelligence (CTI) on actors targeting the organization \cite{automated-threat-hunting-ics}. Threat attribution aids in establishing profiles of threat actors targeting specific sectors \cite{sharma-23} and can inform such sectors on appropriate countermeasures, including threat hunting approaches focusing on techniques employed by relevant threat actors to detect malicious activities before excessive harm is caused \cite{nour-23}. \par
The threat attribution process is nontrivial and is typically performed manually by forensic experts after an incident occurs \cite{fintech-ttp-attribution}. Automatically attributing security incidents as they happen could significantly reduce the effort required to generate actionable intelligence. In addition, such automation could reduce incident detection times and increase the resilience of organizations against Advanced Persistent Threats (APTs). Automatic attribution might also result in additional information for automatically generated CTI feeds. \par
While several proposals have been made to automate threat attribution \cite{automated-threat-hunting-ics,automatic-attribution-mobile,fintech-ttp-attribution}, these systems are not easily compared or combined despite their similar or related goals. Current research tackles threat attribution as a complex monolithic problem or even as part of another problem, hence hindering the adaptation or reuse of components. This paper argues for an alternative and proposes a modular approach that subdivides the attribution problem into subproblems that are addressed separately, after which results can be aggregated using opinion pools. Moreover, we propose an aggregation method for the modular architecture that forms pairs of attributors based on different indicator types. \par
Our main contribution is demonstrating that threat attribution does not need to be tackled as a monolithic problem. On the contrary, we suggest tackling it as a modular problem, and show how different modules may cooperate using opinion pools. We argue that the modular approach improves the tractability of the threat attribution problem, increases the usability of threat attribution solutions, and aids the interpretability of threat attribution outcomes. Moreover, we demonstrate that these improvements can be implemented without adversely affecting the classification performance. \par
To this end, related work on threat attribution and opinion pools is discussed in Section~\ref{sec:background}. Thereafter, the approach is formalized and its envisioned benefits are discussed in Section~\ref{sec:approach}. In order to examine the viability of the proposed approach, an experimental validation is conducted in Section~\ref{sec:experiment} to compare the classifying capabilities of the proposed approach with baselines. The section also briefly addresses the computational complexity and illustrates interpretability through an example. Finally, limitations and potential enhancements of the approach are discussed in Section~\ref{sec:discussion}, and Section~\ref{sec:conclusion} concludes this work.

\section{Background and related work}
\label{sec:background}
In order to provide insights into current research relating to threat attribution and opinion pools, the most relevant papers are discussed below. First, Section~\ref{sec:threat_attribution} introduces the concept of threat intelligence sharing, and describes the threat attribution process in more detail by highlighting the characteristics of the threat attribution process and discussing several existing automated methods for threat attribution and related problems. Thereafter, Section~\ref{sec:opinion_pools} provides a brief summary on opinion pools, which are used as building blocks for the proposed aggregation function.
\subsection{Cyber Threat Attribution and related topics}
\label{sec:threat_attribution}
\subsubsection*{Cyber threat intelligence and sharing}
In a survey, Tounsi and Rais define \textit{Cyber Threat Intelligence (CTI)} as evidence-based knowledge that can inform decisions concerning threats \cite{tounsi-18}. CTI on a high level may describe the identity of threat actors along with their goals, but may also describe low-level observations such as Indicators of Compromise (IoCs) \cite{bromander-16}. Standardized formats exist for CTI, such as MITRE ATT\&CK for describing the behavior of threat actors \cite{mitre-attack}. Moreover, formats such as the Structured Threat Information Expression (STIX) language exist to describe relations between objects such as IoCs, techniques, campaigns, and actors \cite{stix}, along with protocols such as Trusted Automated Exchange of Intelligence Information (TAXII) to share them \cite{taxii} and open-source implementations such as Malware Information Sharing Platform (MISP) \cite{misp}.
\subsubsection*{Threat attribution frameworks}
\textit{Cyber threat attribution} is the process according to which the actors responsible for cyber threats are determined. Rid and Buchanan proposed the Q model, which is designed to guide investigators through the attribution process \cite{attributing-cyber-attacks}. They emphasize uncertainty in the attribution process exist on three levels: tactical, operational, and strategical. Questions regarding how incidents occurred, what happened, and why they were caused play a central role in this model. They also emphasize the importance of how communication about the attribution process should take place. A related contribution is a book by Timo Steffens describing several attribution methods utilizing different information sources \cite{Steffens_2020}. \par
Pahi and Skopik proposed the Cyber Attribution Model (CAM) to guide forensic investigators through the attribution process while avoiding the pursuit of false-flags \cite{cyber-attribution-model}. \textit{False-flags} are indicators left by a threat actor to evade attribution or to evoke blame upon another entity. According to the model, the attribution process is best understood by asking what infrastructure was used to carry out the operation, what capabilities the attacker must have had, and what the motivation of the attacker could have been. Sociopolitical and technical contextual indicators must be considered to detect false-flags. They have demonstrated the value of the model by applying it to the 2015 TV5Monde hack. \par
Skopik and Pahi later investigated what classes of information are considered most trustworthy for attributing an incident to a threat actor \cite{under-false-flag}. The trustworthiness was determined by surveying several domain experts. They found that the general Tactics, Techniques, and Procedures (TTPs), cloud services, Command \& Control (C2) infrastructure, and DNS patterns used are the most trustworthy features for attribution. On the other hand, traces on the dark web consistent with technical artifacts, phishing attempts, and local malware and their properties were considered the least trustworthy.
\subsubsection*{Automated threat attribution approaches}
While attribution models are well studied, automated approaches to establish attribution are sparse in academic literature. Noor \textit{et al.} investigated the usage of Tactics and Techniques (TTs) as high-level attack patterns for cyber threat attribution \cite{fintech-ttp-attribution}. In order to train a machine learning model, the TTs used by known threat actors are collected in a correlation matrix as zeros and ones. The authors also proposed a method for extracting TTs from unstructured CTI documents to fill the matrix more fully, instead of drawing from the ATT\&CK knowledge base. They compared various machine learning models, and their results suggest that a neural network model is the most promising for this task. \par
Recent efforts to automate the threat attribution process were made by Kim \textit{et al.} \cite{automatic-attribution-mobile}. In their work, they used the ATT\&CK for Mobile taxonomy to perform attribution and extracted threat intelligence concerning the used TTs from the ATT\&CK Matrix. In order to classify a sample, it was executed in a sandbox to extract the TTs associated with that malware sample. The techniques used for each tactic were then expressed as a vector. The similarity between the observed vectors and vectors from known threat intelligence was measured to classify the samples. They utilized pairs of IOCs instead of individual IoCs to improve precision and avoid following false-flags. \par
A related contribution was made by Arafune \textit{et al.} \cite{automated-threat-hunting-ics}. They aim to predict the future steps of the attacker through threat attribution using adversarial TTs. Their approach filters out malicious traffic using a Network Intrusion Detection System (NIDS) and extracts TTs from the selected traffic. In addition, they used a support vector machine to predict future steps of the assumed multistep attack. Their work shows how threat attribution can be used to improve detection methods.
\subsubsection*{Attribution in other domains}
Attribution also serves a variety of purposes outside the cybersecurity domain. Du \textit{et al.} describe several applications of machine learning, which they refer to as Artificial Intelligence (AI), for various fingerprinting techniques \cite{du-20}. These include malware attribution, digital message authorship attribution, and photo hardware origin attribution. While these attribution problems can be considered more restricted than the threat attribution problem, they share a similar goal.
\subsubsection*{Gap analysis}
The automated approaches for threat attribution we identified all view attribution as a monolithic problem, and the solutions they propose are not easily interoperable or interchangeable. Moreover, black-box machine learning models like the aforementioned neural networks can yield predictions that are difficult to interpret \cite{fan-21} for forensic experts, who might instead require convincing evidence for prosecution. The proposed modular approach attempts to overcome these problems while increasing the tractability of the threat attribution problem. Concretely, the approach elucidates how to perform automatic attribution based on multiple evidence obtained from distinct attribution approaches. The proposed approach is interoperable with existing automated solutions such as those described earlier for threat attribution \cite{fintech-ttp-attribution, automatic-attribution-mobile, automated-threat-hunting-ics}, and may also be applicable to other types of fingerprinting such as authorship attribution \cite{du-20}. Moreover, the modular approach draws from the CAM \cite{cyber-attribution-model}, related models \cite{attributing-cyber-attacks}, and the pairing of IoC \cite{automatic-attribution-mobile}, and combines these on an architectural level. Research on the trustworthiness of specific indicators for attribution \cite{under-false-flag} can support the operationalization of the proposed framework.
\subsection{Opinion Pools}
\label{sec:opinion_pools}
Opinion pools were first introduced by Stone in 1961 as a function to make a joint decision based on the opinions of several individuals \cite{stone-61}. Opinions can be modelled as a function describing probabilities for each possible decision. Examples of such functions are probability mass and density functions. A \textit{Probability Mass Function (PMF)} is a function that describes the possible values of a discrete random variable using probabilities, and a \textit{Probability Density Function (PDF)} describes the same for continuous random variables. \par
Koliander \textit{et al.} recently conducted a survey on the fusion of PDFs using opinion pools \cite{fusion-pdf}. Formally, \textit{Opinion pools} are functions that combine $K \geq 1$ PDFs into a single aggregate PDF. These pooling functions can typically also be applied to PMFs. Various opinion pools exist, and they differ in terms of the desirable axioms they satisfy. In the remainder of this work, only the linear opinion pool and the logarithmic opinion pool are considered, although also other pooling functions exist. In particular, the holder opinion pool is noteworthy as it is parameterizable and can be considered a generalization of both the linear opinion pool ($\alpha=1$) and the logarithmic opinion pool ($\alpha=0$). \par
The linear opinion pool is computed as an arithmetic average, as shown in Equation~\ref{eq:linear_pool} where the function $g$ produces the aggregate PDF and $q_k(\theta)$ describes an input PDF. The logarithmic opinion pool is computed as the geometric average, as shown in Equation~\ref{eq:logarithmic_pool} where $c$ is a normalization factor. Both pools can be weighted using a factor $w_k$ for each input PDF. For the sake of simplicity, equal weights are used in the remainder of this work.
\begin{equation}
    \label{eq:linear_pool}
    g_{linear}[q_1, \hdots, q_K](\theta) = \sum_{k=1}^{K} w_k q_k(\theta)
\end{equation}
\begin{equation}
    \label{eq:logarithmic_pool}
    g_{logarithmic}[q_1, \hdots, q_K](\theta) = c \prod_{k=1}^{K} q_k(\theta)^{w_k}
\end{equation}

\section{Proposed approach}
\label{sec:approach}The aim of this work is to propose an architecture that can be used to attribute incidents to known threat actors automatically and to support forensic experts in the attribution process by suggesting actors who are likely responsible for incidents. Before detailing this architecture, relevant definitions are provided. \par
According to the NIST glossary, any observable occurrence in a network or system is an \textit{event} \cite{nist-glossary}. We define an \textit{incident} as a set of related events from which several Indicators of Compromise can be derived. This definition is in-line with a definition from the NIST glossary that defines an incident as an \enquote{anomalous or unexpected event, set of events, condition, or situation at any time during the life cycle of a project, product, service, or system} \cite{nist-glossary}. \textit{Indicators of Compromise (IoCs)} are evidence of breach and may also be used for attribution to threat actors. A \textit{threat actor} is a malicious entity whose actions contribute to the cause of an incident. We define an \textit{attributor} as a module that performs attribution of incidents to threat actors. While attributors may directly use IoCs for attribution, they can also utilize extracted features and enrich data. For example, from domain names entropy measures, registrar information, and resolved IP addresses may be derived, which can all inform attribution. \par
When devising a system that performs threat attribution, it is important to account for the needs of various stakeholders, such as forensic experts, system developers, and researchers. These needs go beyond the functional requirement that the system attributes incidents to threat actors. Non-functional requirements for automated approaches have been analyzed by Habibullah \textit{et al.} \cite{habibullah-23}. A goal of the proposed approach is to account for some of these requirements. Several non-functional requirements which the proposed approach aims to satisfy, are discussed in detail in Section~\ref{sec:benefits}. \par
The inspiration for the proposal draws strongly from the fields of software architecture and software design, where non-functional requirements play an important role and design patterns are used to satisfy these requirements better. The most prominently applied design patterns are the Composite and Strategy patterns \cite{design-patterns}. Moreover, the work by Kim \textit{et al.} \cite{automatic-attribution-mobile} inspired the proposed Pairing Aggregator. \par
\begin{figure}[tbp]
    \centering
    \includegraphics[width=1.0\linewidth]{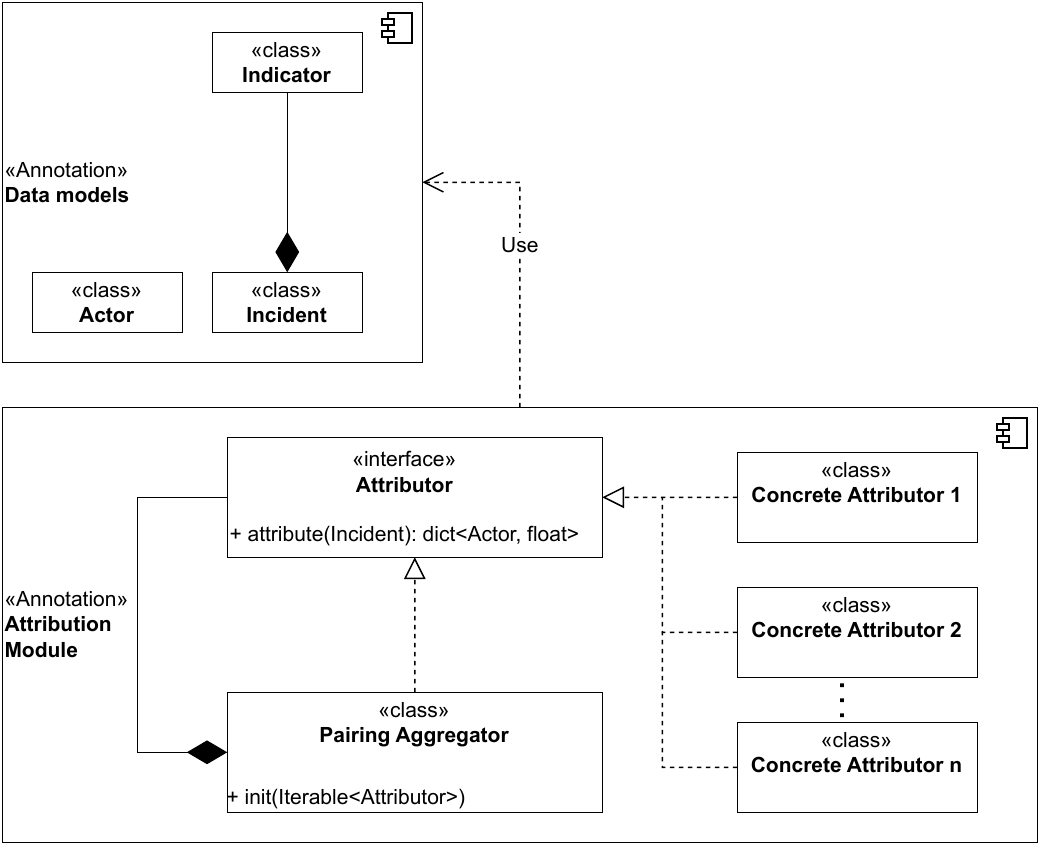}
    \caption{Class diagram describing the proposed modular architecture for threat attribution.}
    \label{fig:attribution_class_diagram}
\end{figure}

Contrary to existing solutions, the newly proposed solution is highly modular and allows for composing attributors at runtime. Attributors can be used interchangeably as they implement a common interface for attribution operating on the same data model, which consists of a single incident composed of multiple indicators as input. Thus, when an improved attributor is implemented, it can easily replace the previous one. The output of the attributor interface consists of a PMF that describes the probability of attribution to known threat actors. An aggregator may combine predictions from different attributors into a single prediction that is better than each PMF produced by the individual attributors, while implementing the same interface. The aggregator can be composed at runtime using any collection of attributors that implement the same interface. Figure~\ref{fig:attribution_class_diagram} shows a class diagram for the proposed architecture. Section~\ref{sec:interpretation} contains more details regarding the input-output of the mentioned interfaces and their interplay.
\par
The individual attributors might correspond to the infrastructure, capabilities, and motivation as described in the CAM \cite{cyber-attribution-model}. Moreover, recursion may, for example, be applied to split the problem of performing attribution based on the infrastructure into the problems of performing attribution leveraging IP addresses and domain names separately. \par
\begin{figure}[tbp]
    \centering
    \includegraphics[width=1.0\linewidth]{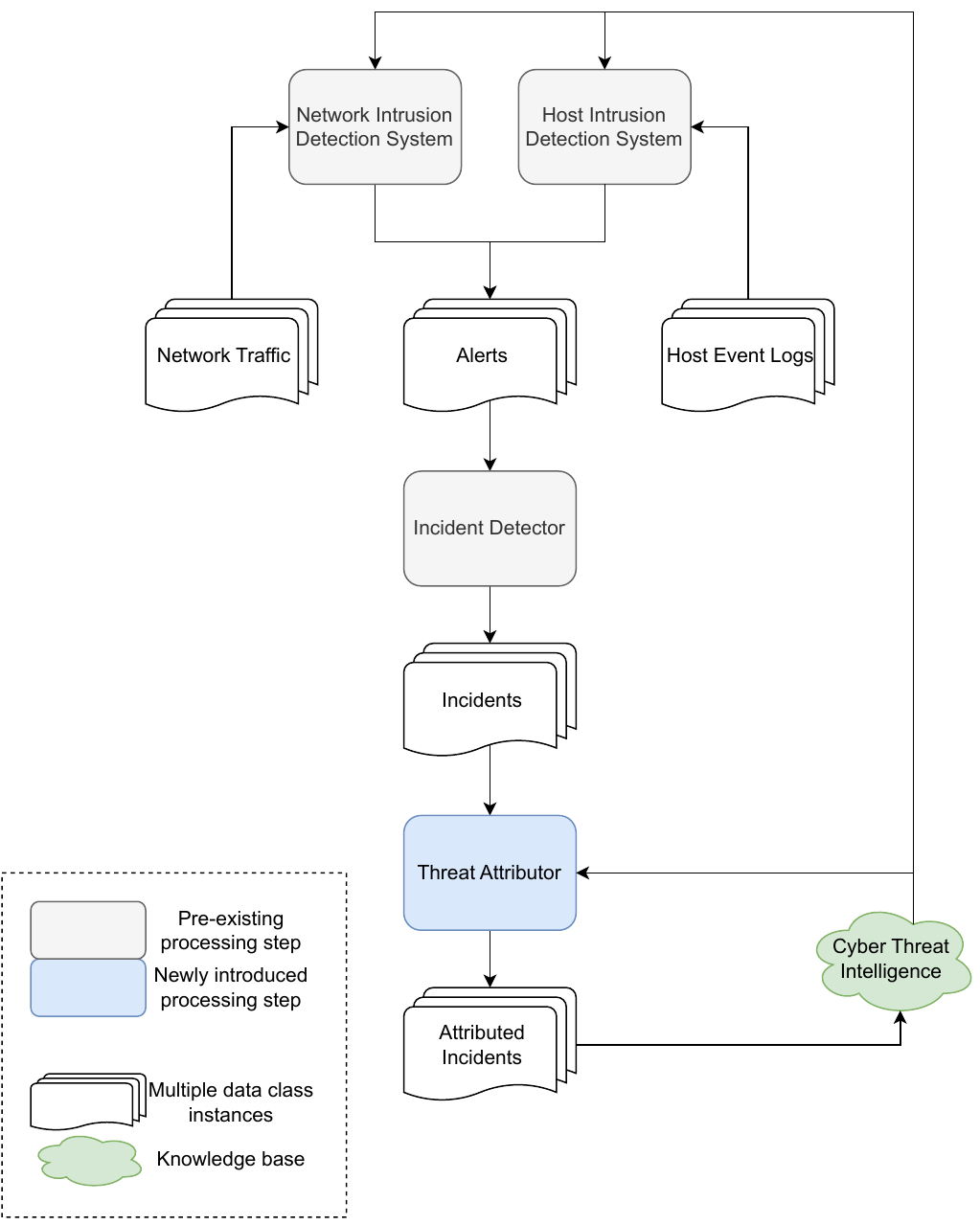}
    \caption{Workflow how the proposed threat attribution approach is related to other areas of research such as intrusion detection systems.}
    \label{fig:attribution_workflow_diagram}
\end{figure}

Similar to related work, the proposal also allows for automation. Figure~\ref{fig:attribution_workflow_diagram} shows what such an integration could look like. Utilizing available CTI, intrusion detection systems can be improved with a focus on the TTPs used by the threat actors targeting the organization, industry, or country. Alerts can then be correlated and aggregated into incidents. All the indicators that can be derived from the incident can then be fed into the threat attribution module, generating additional threat intelligence. \par
\par
In order to further define the proposal, a more formal description is provided. First, we define the set of known threat actors $T$, a subset of real numbers $\mathbb{R}_{(0,1)} = \{ x \ | \ x \in \mathbb{R} \land 0 \leq x \leq 1\}$, and the set of PMFs $Q = \{ q \ | \ q: T \rightarrow \mathbb{R}_{(0,1)} \land \sum_{t \in T} q(t) = 1\}$. Consider a set of attributors $A$ and an incident with indicators $I$. Each attributor $a \in A$ can be considered a function taking a subset of indicators $I' \subseteq I$ as input and mapping this to a valid PMF $q_a \in Q$ indicating the empirical probabilities of attribution to known threat actors $T$. The function implemented by each concrete attributor is the following: $A: \mathcal{P}(I) \rightarrow Q$\footnote{$\mathcal{P}$ is used to denote the power set.}. An aggregator $\mathcal{A}$ is composed of a set of attributors $A' \subseteq A$ and maps their output PMFs $Q' \subseteq Q$ to an aggregate output PMF: $\mathcal{A}: \mathcal{P}(Q) \rightarrow Q$. \par
Ideally, the aggregator preserves nuances present in the input PMFs, indicating uncertainty in the attribution. Therefore, a majority vote, for example, could be considered unsuitable. Moreover, if the input PMFs are more contradictory, the output PMF should express a larger degree of uncertainty. This property is satisfied by the linear opinion pool. Another desirable property for the aggregator is that it separates the likely and unlikely outcomes, such that automatic attribution might take place using a clear threshold. This property is satisfied by the logarithmic opinion pool. \cite{fusion-pdf}\par
In order to aggregate the output of the individual attributors, a Pairing Aggregator is proposed with the aim of satisfying these characteristics and increasing the resiliency against false-flags. The underlying idea for this aggregator is that the attribution precision can be improved by inspecting multiple indicator types simultaneously. First, the Pairing Aggregator forms pairs of attributors and combines their PMFs using a logarithmic opinion pool, after which the intermediate PMFs are combined using a linear opinion pool to derive the aggregate PMF. More formally, consider the set of input PMFs $Q'$ resulting from concrete attributors $A'$. The pairing attributor forms pairs $P$ according to Equation~\ref{eq:pairs}. It then produces intermediate PMFs $R$ according to Equation~\ref{eq:intermediate_results} using logarithmic opinion pools, after which the final aggregate PMF is computed according to Equation~\ref{eq:final_result} using a linear opinion pool.
\begin{equation}
    \label{eq:pairs}
    P = \{ (q_1, q_2) \ | \ q_1, q_2 \in Q' \land q_1 \neq q_2\}
\end{equation}
\begin{equation}
    \label{eq:intermediate_results}
    R = \{ g_{logarithmic}[x, y] \ | \ (x, y) \in P\}
\end{equation}
\begin{equation}
    \label{eq:final_result}
    g_{linear}[r \ | \ r \in R]
\end{equation}
\subsection{Benefits}
\label{sec:benefits}
Three main benefits offered by the proposed approach have been identified. These pertain to the tractability of the threat attribution problem, to the usability of conducted research and developed solutions, and to the interpretability of the results. The modular architecture is approach-agnostic in the sense that related work can be integrated into the proposed architecture as concrete attributors. Since it is challenging to evaluate these benefits numerically, we provide arguments for why the modular approach offers advancements in these three areas. Complementing these arguments, the experiment described in Section~\ref{sec:experiment} aims to compare classification capabilities with baselines and provides an example to highlight interpretability.
\subsubsection{Tractability}
Purao formulated that research problems should be tractable, implying their scope should be small enough to solve \cite{research-problems}. One may argue that threat attribution on its own is intractable so far. Rid and Buchanan also highlight the complexity of the threat attribution process and even name it \enquote{an art} \cite{attributing-cyber-attacks}. Hence, the need for decomposing the problem into more tractable problems exists. \par
Similar to the divide and conquer approach from algorithm design \cite{divide-and-conquer}, the proposal is to split the problem of how to attribute incidents into smaller problems that are easier to solve. Instead of addressing the complete problem, researchers can, for example, focus on problems such as how to attribute an incident based on observed domain names. These smaller problems can be easier to solve, and can later be used to address the larger problem by combining the solutions to the subproblems. The decomposition could be derived from threat attribution models such as the CAM \cite{cyber-attribution-model}.
\subsubsection{Usability}
Moreover, researchers might find themselves working on distinct subproblems whose solutions can cooperate by leveraging the modular approach, as opposed to competing for the best solution to the overall problem. Research on software reuse indicates that smaller modules have the largest chance of being completely reused \cite{software-reuse}. Enabling reuse should be considered a goal in research to enable cooperation and support future research. \par
From an operational perspective, the concrete attributors from the modular approach can be considered building blocks for other components which can be constructed by means of composition. Reuse of software is encouraged since additional modules can easily be added to extend an already existing system. Hence, the modular approach also enables interoperability between existing threat attribution solutions. Moreover, maintainability benefits from modularity—replacing a single module does not introduce the need to decommission the whole system since a module can be effortlessly replaced if it implements the same interface. Another benefit in terms of usability is that the modular approach allows for the parallelization of concrete attributors. This parallelization is even supported across different devices due to the portability of the decoupled modules.
\subsubsection{Interpretability}
The modular approach has the additional benefit that it naturally enables interpretability and provides some sort of provenance. Explainability of machine learning has received increased attention in recent years \cite{explainable-ai}, stressing the importance of interpretability. The ability to understand why an automated attribution system suggests a certain threat actor is crucial if forensic experts intend to follow up on automated outcomes and build a case for prosecution to hold responsible threat actors accountable for their actions. If, instead, interpretability were lacking, the technical model did not help at all and the experts would have to start from scratch despite an attempt at automation. It should also be noted that PMFs are useful for interpreting attribution outcomes in the case of incidents where multiple actors are involved, such as the case where a threat actor distributes malware through a service provided by another actor \cite{huang-17}, because a PMF may suggest several actors. \par
The modular approach aids interpretability due to its transparent structure. If each module performs a fixed small task and intermediate results for the attribution process are available, it is naturally clear which indicators contribute to the attribution outcome of an incident. For example, if a module that performs attribution using domain names and a module using TTPs both agree with the final attribution, it is clear that these features contribute to this result. Section~\ref{sec:interpretation} further elaborates on the interpretability using example output.

\section{Experimental validation}
\label{sec:experiment}
In order to assess the potential value of a modular approach, an experiment is conducted to compare the modular approach with monolithic approaches. The goal of the experiment is to compare the classification capabilities and runtimes of the modular approach with the monolithic alternatives. First, the experiment design is described in Section~\ref{sec:experiment_design}. Thereafter, a detailed description of the used dataset is provided in Section~\ref{sec:simulated_dataset_generation}. The results of the experiment are presented and discussed in Section~\ref{sec:results}. The computational complexity of the proposed solution is assessed in Section~\ref{sec:complexity} and a practical example of the output is interpreted in Section~\ref{sec:interpretation}. The source code used to conduct the experiments is made publicly available to enable reproducibility.
\subsection{Experiment design}
\label{sec:experiment_design}
To the best of our knowledge, no good public dataset exists for studying threat attribution \footnote{The sources queried include IEEE DataPort and Kaggle.}. Prior work has either scraped the internet \cite{automated-threat-hunting-ics,fintech-ttp-attribution}, or used labeled sandboxed malware executions \cite{automatic-attribution-mobile}. The representativeness of each of these methods is imperfect, and a perfectly representative dataset is unachievable due to the adversarial nature of the domain. Since the experiment aims to validate the architecture instead of concrete attributors, a simple artificial dataset with $8$ non-stationary features is generated with the aim of capturing the essence of how cybersecurity incidents occur in practice. Each feature can be considered to describe a group of related facts that can be derived from an incident. A total of $392\,577$ incidents is derived from $128$ generated threat actor profiles. The features of an incident are sampled according to the profile of the responsible threat actor, which changes over time. Moreover, actors may try to evoke blame on others by leaving false-flags that are associated with a different threat actor. A detailed description of the procedure to generate this dataset is given in Section~\ref{sec:simulated_dataset_generation}, which also includes several descriptive statistics on the dataset. The dataset captures the various complexities encountered in the threat attribution process sufficiently well to allow for comparing classifier performance without the intention of drawing conclusions relating to the performance of a specific classifier in practice. \par
The experiment consequently compares the modular approach with a complex monolithic solution. The eXtreme Gradient Boosting (XGBoost) classifier \cite{xgboost} was chosen as a baseline to represent the monolithic approaches, since XGBoost is known to perform well on a variety of tasks and can outperform more complex deep neural networks \cite{xgboost-performance}. This classifier was implemented\footnote{Where unspecified default parameters were used in combination with \textit{objective='multi\_softmax'} for \url{https://xgboost.readthedocs.io/en/stable/parameter.html}} with $1\,000$ estimators. \par
For the modular approach, one concrete attributor is implemented for each of the features. The concrete attributors are implemented\footnote{Default parameters were used in combination with \textit{kernel='linear'} and \textit{probability=True} for \url{https://scikit-learn.org/1.2/modules/generated/sklearn.svm.SVC.html}} as Linear Support Vector Machines (SVMs) operating only on a single feature. Consequently, the Pairing Aggregator described in Section~\ref{sec:approach} is used to combine the individual results into a final PMF. The proposed aggregator is also compared with the standalone linear and logarithmic opinion pools. In addition to XGBoost, a Linear SVM is also implemented operating on all features as a representative for the monolithic approaches. \par
Since the attribution framework is meant to support forensic experts and should suggest the most probable responsible threat actors, the top $k$-accuracy score is used to evaluate the various alternatives. The top $k$-accuracy for a sample is computed as the smallest $k$ for which the responsible actor is among the top $k$ most probable outcomes of the resulting PMF. In addition, we compute the precision and recall, which are defined respectively as the fraction of correctly attributed incidents among the attributed incidents and the fraction of correctly attributed incidents among the incidents that could be attributed correctly with a sufficiently high threshold. Using these metrics, the micro-averaged Precision-Recall (PR) curve is shown for the various alternatives, and their optimal F-measures are computed.
\subsection{Simulated dataset generation}
\label{sec:simulated_dataset_generation}
In order to obtain a suitable dataset, a dataset of incidents caused by threat actors is simulated using $s=100\,000$ time steps. The first $\frac{s}{2}$ time steps are used as training data, while the remaining time steps are used as test data. Each incident in this dataset is abstractly described by $m=8$ numerical features. Categorical features such as those describing the use of specific TTPs can also be represented through, for example, one-hot encoding. First, profiles are created for $t=128$ threat actors. Each threat actor $i$, with $0 \leq i < t$, has an activity rate $a_i \sim \mathcal{U}(0.0001, 0.1)$\footnote{$\mathcal{U}$ is used to depict a uniform distribution.}, which represents the probability that the threat actor causes an incident at the given time step. Each threat actor $i$, with $0 \leq i < t$, commences their activities at $s_{i\_start} \sim \mathcal{U}(0, 0.4s)$ and ends their activities at $s_{i\_end} \sim \mathcal{U}(0.6s, s)$. This corresponds to the expected behavior that new actors can occur over time and others cease to operate as they are, for example, prosecuted. Moreover, each actor $i$, with $0 \leq i < t$, has feature means $x_{i\_j\_\mu} \sim \mathcal{U}(-1, 1)$ and features standard deviations $x_{i\_j\_\sigma} \sim \mathcal{U}(0, \frac{1}{t})$ for each feature $x_{i\_j}$ with $0 \leq i < t$ and $0 \leq j < m$. \par
At each time step and for each threat actor $i$, with $0 \leq i < t$, an incident is generated with probability $a_i$. The features representing the incident are sampled like $x_j \sim \mathcal{N}(x_{i\_j\_\mu}, x_{i\_j\_\sigma})$\footnote{$\mathcal{N}$ is used to depict a normal distribution.} for each feature $x_{i\_j}$ with $0 \leq i < t$ and $0 \leq j < m$. For each of the features, with probability $0.4$, the feature value in the test data is replaced by a false flag randomly chosen from the feature values of a different actor in the already generated training data. Moreover, to simulate a degree of non-stationarity in the behavior of the threat actors, at each time step, the feature means shift with $x_{i\_j\_\mu\_shift} \sim \mathcal{N}(0, 0.01)$ and the feature standard deviations change with $x_{i\_j\_\sigma\_shift} \sim \mathcal{N}(0, 0.01)$ for each feature $x_{i\_j}$ with $0 \leq i < t$ and $0 \leq j < m$. Similarly, the activity level $a_i$ of each threat actor $i$, with $0 \leq i < t$, changes every time step with $a_{i\_shift} \sim \mathcal{N}(0, 0.01)$ The resulting dataset includes both prior probability shift and concept shift, which are different forms of dataset shift \cite{moreno-torres-12}. Hence, the resulting dataset describes a nontrivial problem for machine learning approaches. \par
\begin{figure}[tbp]
    \centering
    \includegraphics[width=1.0\linewidth]{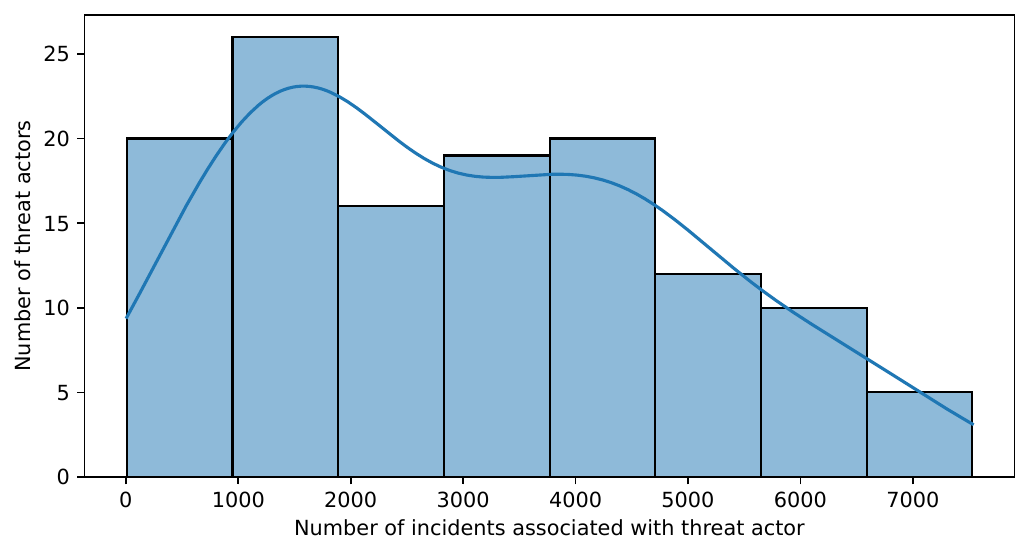}
    \caption{Combined histogram and kernel density estimation, depicting the distribution of the number of incidents associated with a threat actor.}
    \label{fig:threat_actor_hist}
\end{figure}
\begin{figure}[tbp]
    \centering
    \includegraphics[width=1.0\linewidth]{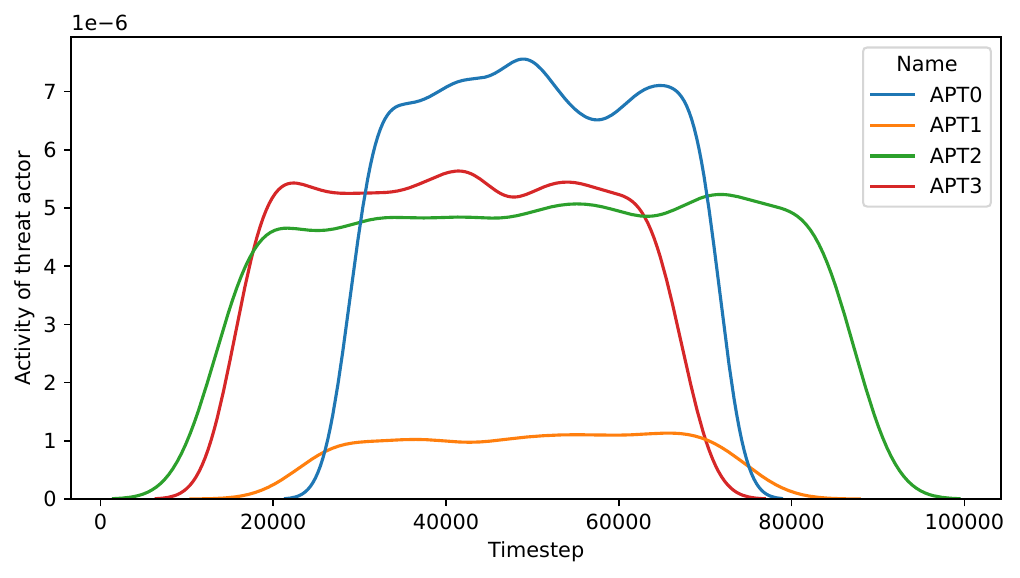}
    \caption{Kernel Density Estimations (KDE) for the number of incidents associated with the first four threat actors over time.}
    \label{fig:threat_actor_activity}
\end{figure}

After generation, the final dataset consists of $392\,577$ samples with $8$ distinct features belonging to $128$ classes representing different threat actors. To characterize the dataset, we highlight several aggregate statistics. Figure~\ref{fig:threat_actor_hist} shows the distribution of the number of incidents associated with a threat actor, which is influenced by the different activity levels of various threat actors. The figure shows that there is an imbalance in the sense that most threat actors have a lower number of incidents associated with them, whereas only few threat actors have a high number of incidents associated with them.
Figure~\ref{fig:threat_actor_activity} shows how the activity levels of the first four threat actors vary amongst each other and over time.
\subsection{Results}
\label{sec:results}
\begin{figure}[tbp]
    \centering
    \includegraphics[width=1.0\linewidth]{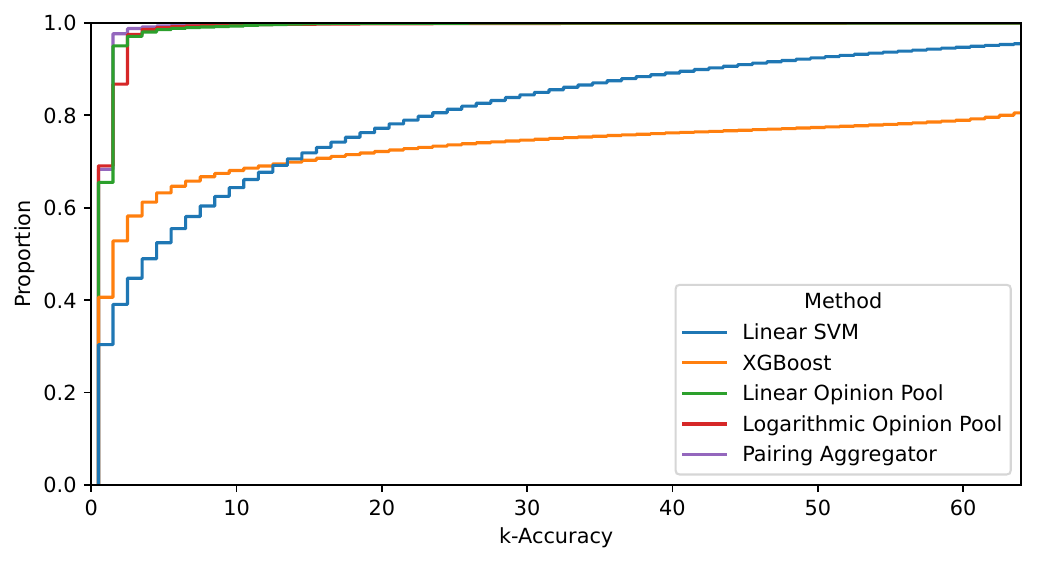}
    \caption{Cumulative distribution plots depicting the proportion of samples where the correct actor is among the top $k$ most probable outcomes for the various alternatives.}
    \label{fig:experiment_k_accuracy}
\end{figure}

\begin{figure}[tbp]
    \centering
    \includegraphics[width=1.0\linewidth]{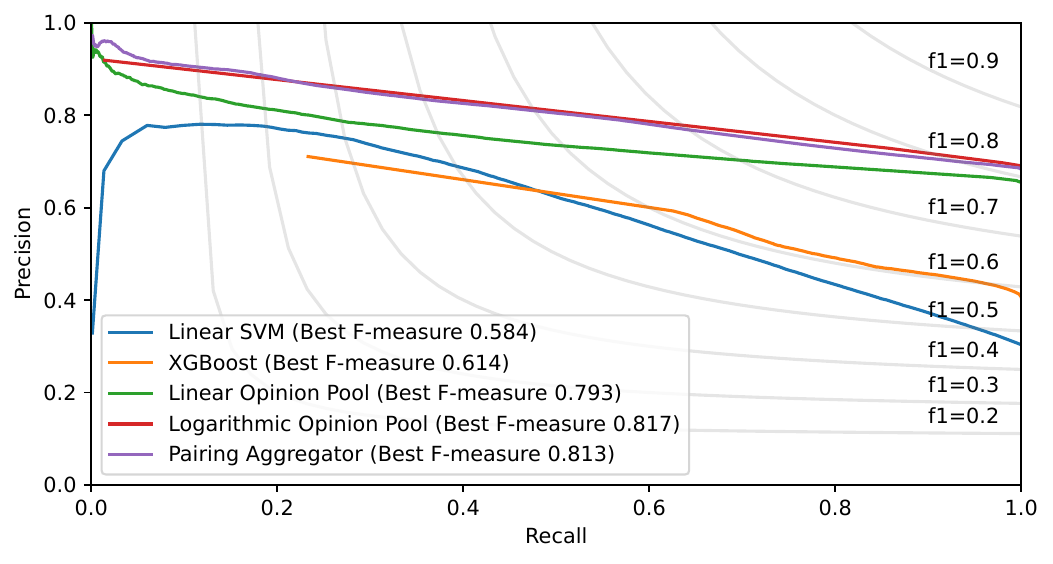}
    \caption{Precision-Recall curves for the various alternatives and their corresponding optimal F-measures.}
    \label{fig:experiment_precision_recall}
\end{figure}

The results of the conducted experiment are shown in two figures. Figure~\ref{fig:experiment_k_accuracy} shows a cumulative distribution plot depicting the proportion of samples where the correct actor is among the top $k$ most probable outcomes. A steep incline on the left of the figure indicates the best performance. Figure~\ref{fig:experiment_precision_recall} shows precision-recall curves for the various attributors and their corresponding optimal F-measures. The most desirable F-measures are located at the top-right of the figure.
\par
The curves corresponding to the opinion pools in Figure~\ref{fig:experiment_k_accuracy} are more closely located towards the top left of the figure compared to the monolithic alternatives. This implies that the list of most probable threat actors best corresponds to the actual responsible threat actor for the modular approach. Hence, the modular approach appears to provide the most useful results for forensic experts who continue an investigation using the aggregate PMF. While XGBoost has more samples with a low $k$-accuracy than the linear SVM, it has fewer samples for which the $k$-accuracy indicates the responsible threat actor is among the top half of most probable threat actors according to the prediction. Nevertheless, the ensemble of the simple models in the modular architecture achieves better $k$-accuracy than the monolithic alternatives. \par
Moreover, Figure~\ref{fig:experiment_precision_recall} shows that the Pairing Aggregator can obtain excellent precision while maintaining acceptable recall levels. Although it obtains a slightly lower F-measure than the logarithmic opinion pool where the F-measure is optimal, the F-measure obtained by the Pairing Aggregator is higher than that of the monolithic solutions such as XGBoost or the single linear SVM. The Pairing Aggregator can also obtain the highest overall precision, suggesting it is more resilient against the presence of false-flags. Although XGBoost outperforms the linear SVM, the alternatives representing the modular approach outperform both monolithic alternatives.
\subsection{Computational complexity}
\label{sec:complexity}
While the runtime of the attribution solution is highly dependent on the chosen underlying machine learning model, we can compare the practical runtime of the modular approach with the monolithic approach where the underlying model is the same or where the performance is comparable. Moreover, we can perform an analysis of the computational complexity where the computational complexity of the underlying model is a parameterized function. \par
Let $\mathcal{O}(r(n, d))$ denote the computational complexity of the underlying machine learning model, where $r$ is a function with $n$ and $d$ as parameters. Parameter $n$ refers to the number of incidents and $d$ refers to the dimensionality resulting from the number of features available to the model. When using the same underlying model, it is both reasonable and conservative to assume $\mathcal{O}(r(n, \hat{d})) \leq \mathcal{O}(r(n, d))$ where $\hat{d} \leq d$ \cite{buczak-16}. The runtime required for pooling $K$ PMFs describing probabilities for $t$ threat actors is bound by $\mathcal{O}(Kt)$ basic operations requiring constant time using a floating-point arithmetic implementation. \par
The proposed pairing aggregator will use at most $d$ models if an attribution module is constructed for every feature. As a result, the logarithmic opinion pool will be applied at most $d^2$ times, followed by a single application of the linear opinion pool for every incident. Therefore, the computational complexity of the Pairing Aggregator is given by $\mathcal{O}(d \cdot r(n, d) + nd^2(dt) + n(d^2t)$, which simplifies to $\mathcal{O}(d \cdot r(n, d) + nd^3t)$. Therefore, the aggregation method is linear with respect to the number of incidents and threat actors. The opinion pools are only applied when making predictions, and the computational complexity during training is therefore limited to $\mathcal{O}(d \cdot r(n, d))$. \par
\begin{figure}[tbp]
    \centering
    \includegraphics[width=1.0\linewidth]{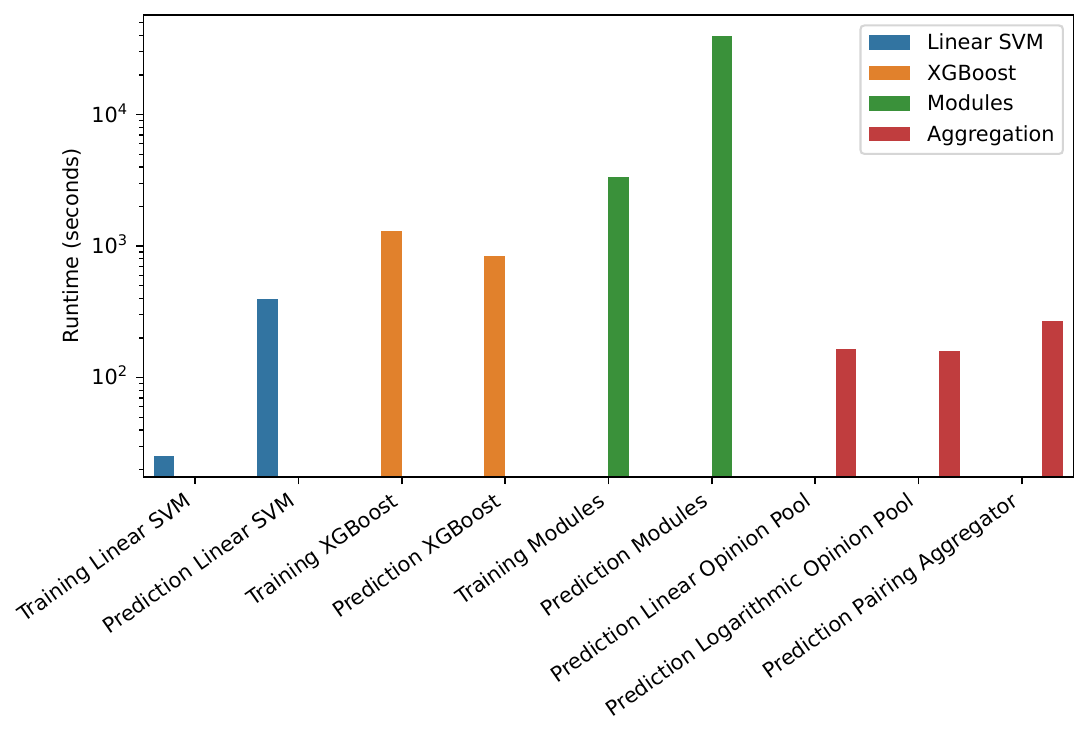}
    \caption{Runtimes of the various alternatives during training and prediction.}
    \label{fig:experiment_runtime}
\end{figure}

While the above-described computational complexity provides a conservative upper bound on the runtime, Figure~\ref{fig:experiment_runtime} shows the practical runtime obtained during the execution of the experiment described in Section~\ref{sec:experiment_design}. These results were gathered on a laptop equipped with an \textsc{Intel i7-11800H} and $32\,\textrm{GB}$ of RAM, with minimal utilization irrelevant to the experiment. All alternatives were executed using a single process. The results indicate that some computational overhead is introduced by the multitude of modules, each running a machine learning model in comparison to the Linear SVM as a baseline, but also indicate that the overhead is not much more than that of more complex models such as XGBoost. Moreover, the runtime required for aggregating predictions using opinion pools is negligible in comparison to the runtime required for gathering predictions from individual modules.
\subsection{Interpretation of the output}
\label{sec:interpretation}
\begin{figure}[tbp]
    \centering
    \includegraphics[width=1.0\linewidth]{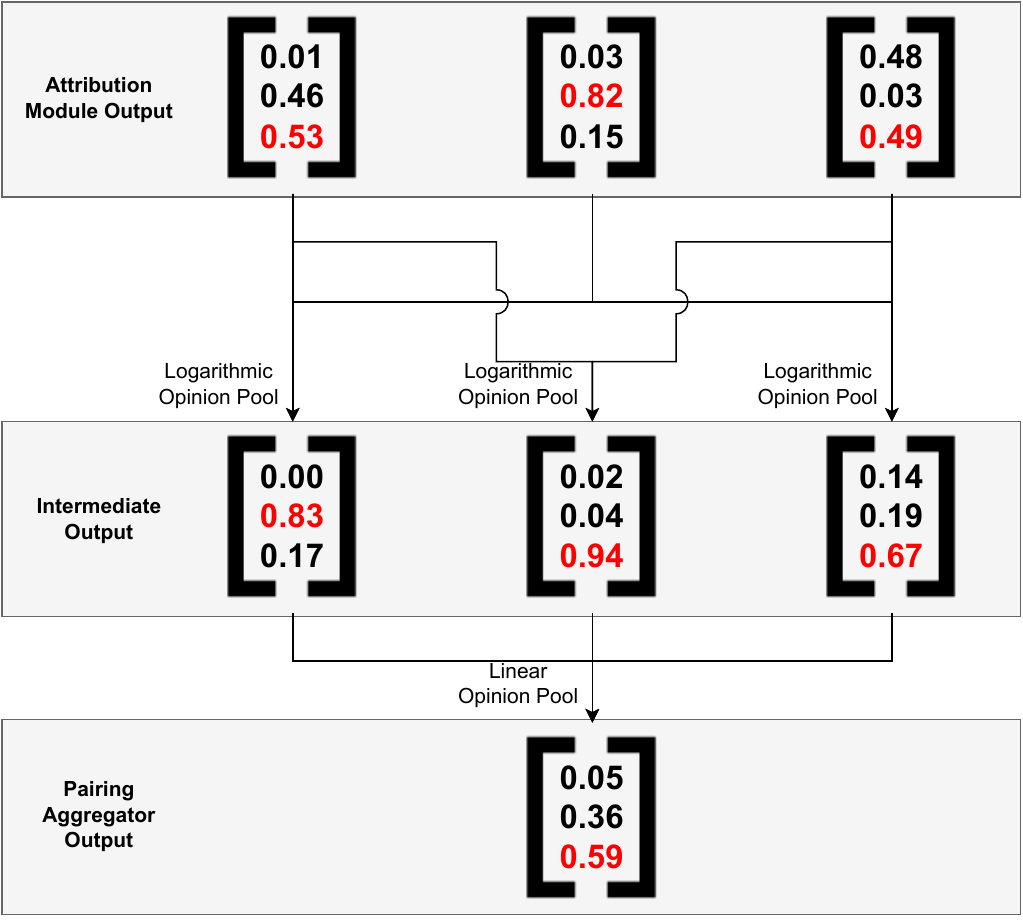}
    \caption{Example of how outputs from individual attribution modules are combined using the Pairing Aggregator.}
    \label{fig:example_interpretability}
\end{figure}

To highlight how the Pairing Aggregator works in practice and how the modular architecture aids interpretability, we use a generated dataset following a method similar to what is described in Section~\ref{sec:simulated_dataset_generation} but with only $3$ threat actors and only $3$ features. Each vector in Figure~\ref{fig:example_interpretability} represents a PMF and depicts the probability of attribution to the threat actor corresponding to the index under consideration. For example, the probability of attribution towards threat actor $t_0$ is $0.01$ in the top-left vector, representing the model output corresponding to feature $f_0$. The top layer of vectors represents the output obtained from the individual attribution modules, and the arrows show how these are combined to produce the intermediate results as described in Section~\ref{sec:approach}. The final output of the Pairing Aggregator is given in the bottom layer.
\par
According to the ground truth corresponding to the example shown in Figure~\ref{fig:example_interpretability}, threat actor $t_2$ is responsible for the corresponding incident. From the output of the individual modules, it becomes clear that features $f_0$ and $f_2$ point in the direction of $t_2$, whereas feature $f_1$ suggests threat actor $t_1$ is responsible instead. Feature $f_1$ can be considered a false-flag in this incident since it misleads forensic experts. Applying the logarithmic opinion pool results in a PMF with high probabilities for the actors on which all input PMFs agree and lower probabilities for the threat actors on which they disagree. A good example of this effect in Figure~\ref{fig:example_interpretability} is given by the combination of the output from the attribution modules operating on features $f_0$ and $f_2$, which agree on the most probable outcome but disagree on the alternatives. The resulting intermediate result has a significant probability of attribution to the threat actor on which the input PMFs agree. After the application of the linear opinion pool, the final outcome of the Pairing Aggregator demonstrates uncertainty since both attribution to actor $t_1$ and $t_2$ are likely. As a result of the interpretation, a forensic expert may further investigate the feature $f_1$ to verify its applicability for attribution of this incident.

\section{Discussion}
Below, a discussion of the conducted work is presented. It is divided into two sections. First, the limitations of the proposed modular approach and the conducted experiments are discussed in Section~\ref{sec:limitations}. Thereafter, potential extensions and enhancements that leverage this work are suggested in Section~\ref{sec:future_work}.
\label{sec:discussion}
\subsection{Limitations}
\label{sec:limitations}
While the proposed modular architecture seems promising, it is important to consider its drawbacks and limitations. First and foremost, the modular approach actually introduces a new problem because the combination of the PMFs of the individual modules may not be straightforward. This does not only apply to the aggregation function itself but also to the optimal structuring of components. Furthermore, in certain scenarios, it might be the case that a monolithic solution achieves better performance, although the conducted experiment demonstrates the modular approach may also outperform monolithic alternatives. \par
In addition, we should consider the possibility that the simulated data used for the experiment is unrepresentative of real-world data. Distributions followed by real features may not follow a normal distribution, and may even have interdependencies with other features. Moreover, all indicators in this simulation had a strong correlation with the responsible threat actor, which may not be the case in practice. Hence, the reported performance scores should only be considered in order to establish a comparison between the alternatives and not to derive conclusions about the performance of concrete attributors in practice.
\subsection{Future work}
\label{sec:future_work}
The approach proposed in this work might be extended or improved in several ways. For an implementation, the modular architecture might be applied recursively. This can, for example, be done by splitting the problem of attribution based on network observables into the problems of attribution based on specific network protocols. Similarly, attribution based on the motivation might be split into attribution based on the type of organization targeted, and the type of actions performed after access was obtained (e.g., information exfiltration, demand for ransom, destructive operations). \par
Moreover, the aggregation function could be further studied and possibly improved. In specific, different aggregation functions might be suitable depending on the types of attributors providing input to the aggregator and the relation between these attributors. The desirable properties for such aggregation functions might be specified in the form of axioms, similar to those discussed in \cite{fusion-pdf}. Weights might also be applied to the pooling functions utilized by the aggregator based on the trustworthiness of the indicator types used for attribution, as investigated in \cite{under-false-flag}. Another important consideration for weighting the opinion pools is that it may be used to reduce the susceptibility to redundant or dependent features. More arbitrary aggregators based on machine learning models might also be possible, although these may decrease the interpretability. \par
Another potential improvement point is the simulation, which is an abstraction of real-world processes. Improvements to the simulation might provide new insights or support testing concrete attributors. Alternatively, the framework might be tested using a real-world dataset or by testing its usability or interpretability in practice in cooperation with forensic experts. This would, however, require concrete attributors to be implemented first.

\section{Conclusion}
\label{sec:conclusion}
The experiment performed in Section~\ref{sec:experiment} demonstrated that the modular approach using opinion pools described in Section~\ref{sec:approach} offers a viable alternative to a monolithic approach. The modular approach may make the threat attribution problem more tractable whilst increasing the usability and interpretability of the threat attribution solution without reducing precision. \par
By making the source code \footnote{\url{https://github.com/Koen1999/modular-threat-attribution}} used to conduct the experiments publicly available, we invite other researchers to experiment with improved simulations, different data, or novel methods for attribution following the proposed architecture. Specifically, the limitations and possible enhancements discussed in Section~\ref{sec:discussion} may be considered. \par
To summarize, the presented modular approach contributes to the effective attribution of digital incidents to threat actors. The architecture allows for the definition of more tractable problems that can contribute to usable and interpretable advancements in the area of threat attribution. On a societal level, attribution may result in prosecution and more effective defense strategies. Overall, the presented work contributes to increased digital security.

\section*{Acknowledgements}
Parts of this work were derived from a prior M.Sc. graduation project \cite{graduation-project}. For their guidance during this project, gratitude is expressed towards Jerry den Hartog, Daniel dos Santos, Elisa Costante, and Alessandro Manzi. Additionally, gratitude is extended to Cristoffer Leite, Luca Allodi, and Emmanuele Zambon for their valuable input while producing this work. Similar gratitude is expressed towards the anonymous reviewers who took the time to provide in-depth feedback. \par
This publication is part of the project CATRIN (with project number NWA.1215.18.003) and the project INTERSECT (with project number NWA.1160.18.301) of the research program Cybersecurity which are (partly) financed by the Dutch Research Council (NWO). For the purpose of Open Access, a CC-BY 4.0 public copyright license is applied to any Author Accepted Manuscript version arising from this submission.

\IEEEtriggeratref{19}

\bibliographystyle{IEEEtran}
\bibliography{IEEEabrv,main.bib}

\end{document}